\documentclass{svjour3}                     
\smartqed  
\usepackage{graphicx}
\usepackage{epsfig}
\usepackage{amsmath}
\usepackage{amssymb}
\usepackage{color}

\begin{document}

\title{Coupled-resonator optical waveguides: $Q$-factor and disorder influence}

\author{Jure Grgi\'c, Enrico Campaioli, S{\o}ren Raza, Paolo Bassi  and Niels Asger Mortensen}


\institute{Jure Grgi\'c, S{\o}ren Raza  and Niels Asger Mortensen\at
            Department of Photonics Engineering,Technical
University of Denmark\\ DK-2800 Kongens Lyngby, Denmark\\
              \email{asger@mailaps.org}           
           \and
           Enrico Campaioli and Paolo Bassi  \at
              Dipartimento di Elettronica, Informatica e Sistemistica, University of Bologna\\
Viale Risorgimento 2, I 40136 Bologna, Italy
}

\date{Received: date / Accepted: date}

\maketitle

\begin{abstract}

Coupled resonator optical waveguides (CROW) can significantly reduce light propagation pulse velocity due to pronounced dispersion properties. A number of interesting applications have been proposed to benefit from such slow-light propagation. Unfortunately, the inevitable presence of disorder, imperfections, and a finite $Q$ value may heavily affect the otherwise attractive properties of CROWs. We show how finite a $Q$ factor limits the maximum attainable group delay time; the group index is limited by $Q$, but equally important the feasible device length is itself also limited by damping resulting from a finite $Q$. Adding the additional effects of disorder to this picture, limitations become even more severe due to destructive interference phenomena, eventually in the form of Anderson localization. Simple analytical considerations demonstrate that the maximum attainable delay time in CROWs is limited by the intrinsic photon lifetime of a single resonator.
\end{abstract}

\section{Introduction}

Slowing down the speed of light enhance phenomena like nonlinearities~\cite{Soljacic:2004}, gain/ab\-sorption~\cite{Sakoda1999},\cite{Grgic2010}, and phase sensitivity~\cite{Soljacic:2002}. It can be also useful for practical applications like delay lines, optical memories, and low threshold lasers~\cite{Scheuer:2005}. CROW structures offer particular ways of slow light guiding, with photons hopping sequentially from one resonator to the next. The concept and proposal of CROWs were introduced in $1999$ by Yariv~\emph{et al.}~\cite{Yariv:1999}. At that point it was obvious that such structures would yield big interest in the optical community due to the potential application possibilities~\cite{Mookherjea:2002,Olivier:2001,Karle:2002,Altug:2004}. In particular, it is clear that near the band edge the group velocity is significantly reduced in the ideal and lossless structure. Unfortunately, a certain level of damping will always be present due to radiation losses, intrinsic losses of materials, and any other mechanism that may dissipate or scatter part of the electromagnetic energy. Any fabrication processes will introduce some variations in the properties of the individual resonators that will serve to dissipate and scatter light propagating in the CROW. Such non-uniformities can indeed seriously affect slow light propagation~\cite{Martinez:2007,Mookherjea:2008,Mookherjea:2002,Mookherjea:2007}, emphasizing the importance of quantifying their effect and unavoidable consequences.

When discussing the potential of slow-light waveguides the emphasis is often on the attainable group index $n_g= c/v_g$ or the group velocity $v_g$. However, in applications involving delay lines and buffers, the group delay $\tau = L/v_g$ is the key parameter. Obviously, the longer the waveguide the longer the group delay! However, this trivial statement implicitly neglects the decay of the pulse as it propagates down the waveguide and eventually the pulse has lost its initial strength and intensity. In this paper, we treat the issues of delay and decay on an equal footing by emphasizing that the decay length $\xi$ serves as an effective cut off for $L$ (section 2), leaving us with a maximal attainable group delay $\tau_{\rm max}$ of the order $\xi/v_g$~\cite{Raza:2010,Grgic:2009}.

The manuscript is organized as follows. From coupled-mode theory (section 3), our key observation is that the maximum attainable delay time in CROWs is limited by the intrinsic photon lifetime of a single resonator (section 3.1). The presence of disorder and scattering will further serve to reduce this bound (section 3.2), as shown numerically with the aid of a Green function method (details given in appendix). Finally, conclusions are given (section 4).

\section{Delay versus decay}

To facilitate quantitative predictions of $\tau_{\rm max}$ we imagine a CROW of length $L$ and the transmission $T$ through this segment is then conveniently parameterized by
\begin{equation}\label{eq:T}
T(\omega)=\exp\left(-\frac{\xi(\omega)}{L}\right).
\end{equation}
Obviously, this parametrization excellently represents the exponential decay of the power associated with absorption and other loss mechanisms captured by the finite $Q$ factors of the resonators. Likewise, in the presence of localization, the average transmission also has an exponential distribution with the scale given by the localization length. Thus, the length scale $\xi(\omega)=-L\ln T(\omega)$ captures the combined effects of disorder-induced localization and other loss accounted for by the finite $Q$-factor.

In the context of pulse delay, $\xi$ represents an estimate of the maximal length of the CROW that we could imagine in any practical application. Extending the length $L$ of CROW beyond $\xi$ would effectively suppress the output power, thus jeopardizing any benefits of slowing down a wave package. The maximal group delay is thus a balance between a slow group velocity and a long propagation length, i.e. $\tau_{\rm max}(\omega)= \xi(\omega)/v_g(\omega)$. Since the group velocity is also inversely proportional to the density-of-states we may conveniently rewrite this expression as

\begin{equation}\label{eq:taumax}
\tau_{\rm max}(\omega)=-\pi \rho(\omega) \ln T(\omega).
\end{equation}

The key result of this paper is that $\tau_{\rm max}$ is always limited by the single-resonator photon lifetime, i.e.
\begin{equation}\label{eq:taumax<taup}
\tau_{\rm max}(\omega)\leq \tau_p
\end{equation}
where the equality applies to the situation with absence of disorder.

In the following we outline our account for this result by coupled-mode theory and in particular, we study first the limit of ideal CROWs (but with finite $Q$) and subsequently we address the effects of disorder.
\section{Coupled-mode theory}
\label{sec:CMT}
We consider a chain of coupled optical resonators, where the $j$th resonator, if left isolated from the other resonators, is characterized by a resonant field
\begin{equation}
\mathbf{E}_j(\mathbf{r},t)= \mathbf{E}_j(\mathbf{r})\exp\left[i(\Omega_j+i\Gamma_j/2)t\right],
\end{equation}
where $\Omega_j$ is the resonance frequency and $\Gamma_j$ represents the resonance line width. The energy in the resonator $\big|\mathbf{E}_j(\omega)\big|^2$ then has a Lorentzian frequency distribution corresponding to the density-of-states
\begin{equation}\label{eq:resonance}
\rho_j(\omega)= \frac{1}{\pi}\frac{\Gamma_j/2}{(\omega-\Omega_j)^2+(\Gamma_j/2)^2}.
\end{equation}
The associated quality factor $Q_j=\Omega_j/\Gamma_j$ may conveniently be parametrized as a photon life time $\tau_p=Q/\Omega$.

Next, imagine a chain of coupled resonators of the above kind. We follow the work of Yariv and co-workers~\cite{Yariv:1999} and write the electrical field as a linear combination of the isolated resonator fields, i.e. $\mathbf{E}_j(\mathbf{r})=\sum_j \psi_j \mathbf{E}_j(\mathbf{r})$, where the expansion coefficients are denoted by $\psi$ to emphasize the similarities with the notation in Ref.~\cite{Datta1995,Datta2000} for the associated problem of electrons in a quantum wire. In the regime of weak nearest-neighbor coupling the equations linearize to
\begin{equation}
(\Omega_j+i\Gamma_j/2) \psi_j - \gamma_{j+1,j}\psi_{j+1} - \gamma_{j-1,j}\psi_{j-1}= \omega \psi_j
\end{equation}
which has a form resembling the tight-binding chain in condensed matter physics \textcolor{black} {~\cite{Kittel2005,Datta1995}}. To further emphasize this connection we write the coupled equations in a matrix form, i.e. ${\boldsymbol H}{\boldsymbol \psi}=\omega{\boldsymbol \psi}$, where the 'Hamiltonian' matrix $\boldsymbol H$ has elements
\begin{equation}
H_{lj}= (\Omega_j+i\Gamma_j/2)\delta_{lj} - \gamma_{lj}\delta_{l\pm 1,j}
\end{equation}
with $\gamma_{lj}=\gamma_{jl}^*$ so that the off-diagonal part of $\boldsymbol H$ is Hermitian.

\subsection{The influence of finite $Q$ factor}

In the case where the resonators are all identical and arranged in a fully periodic sequence, we may without loss of generality suppress all indices, thus making further analytical progress possible. The electromagnetic states now form a continuous frequency band and the problem is easily diagonalized by the Ansatz $\psi_{j+1}=\exp(i \kappa a) \psi_j$. The resulting dispersion relation is of the form
\begin{equation}\label{eq:dispersion}
\omega(\kappa)= \Omega\left(1+i \frac{1}{2Q}\right)-2\gamma\cos(\kappa a)
\end{equation}
where $\kappa=\kappa'+i\kappa''$ is the complex valued Bloch wave vector and $a$ is the lattice constant of the periodic arrangement of resonators. Eq.~(\ref{eq:dispersion}) corresponds to the theory by Yariv~\emph{et al.}~\cite{Yariv:1999} to also include
resonators with a finite $Q$-factor.\footnote{To ease the comparison to our previous work in Ref.~\cite{Raza:2010} we note that
$\omega(\kappa)= \Omega\left(1+i \frac{1}{2Q}\right)\left[1 -\tilde\gamma\cos(\kappa a)\right]$ where $\tilde\gamma$ is for simplicity considered real and given by $\tilde\gamma = 2\gamma/\Omega$.}

The group velocity may formally be calculated from the dispersion relation in Eq.~(\ref{eq:dispersion}). We imagine the situation where the CROW is excited by a monochromatic laser with a well-defined frequency of the light. Care should thus be taken that $\omega$ is to be considered real while $\kappa$ may be complex, and the group velocity is then formally given by

\begin{equation}\label{eq:vg}
v_g = \frac{1}{ {\rm Re} \left\{\partial\kappa/\partial\omega \right\}}.
\end{equation}

Along the same lines, we may also calculate the density-of-states from the dispersion relation in Eq.~(\ref{eq:dispersion}). We emphasize, that for the particular case of a one-dimensional chain, the density-of-states is inversely proportional to the group velocity,
\begin{equation}\label{eq:rho_CROW}
\rho(\omega) = \frac{a}{\pi}{\rm Re} \left\{\frac{\partial\kappa}{\partial\omega }\right\}.
\end{equation}

Clearly, a vanishing group velocity will be associated with a diverging density-of-states and vice versa. This also explains why a light pulse can not completely come to stop in a real system, since no real system will exhibit a true singular density-of-states the inevitable presence of absorption, radiation, and imperfections will serve as effective broadening mechanisms. Nevertheless, to illustrate the basic physics, it is common to consider lossless structures and only focus on the real part of the dispersion properties, since the damping is anyway assumed to be modest. However, in the present context of slow-light propagation the two issues are not easily separable and care must be taken.

\begin{figure}[t!]
\begin{center}
\includegraphics[width=\columnwidth]{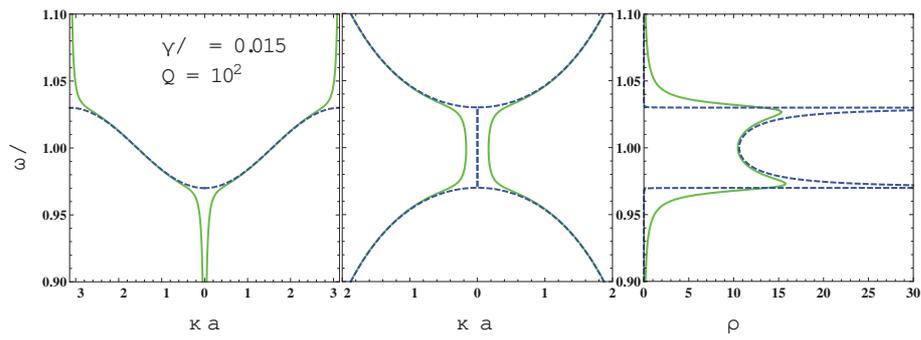}
\end{center}
\caption{Complex dispersion relation for a CROW. Dashed lines are for $Q=\infty$ while solid lines correspond to $Q=10^2$. The left panel shows the frequency $\omega$ versus the real part of the Bloch wave vector $\kappa'$, the middle panel shows the frequency $\omega$ versus the imaginary part of the Bloch wave vector $\kappa''$, and the right panel shows the density-of-states $\rho$ (per resonator).} \label{fig1}
\end{figure}

Figure~\ref{fig1} illustrates the contrast between a CROW made from lossless resonators (dashed lines) and finite-$Q$ resonators (solid lines), respectively.
The left panel illustrates the usual cosine-band dispersion relation, i.e. the relation between the frequency
$\omega$ and the real part $\kappa'$ of the complex-valued Bloch
wave vector $\kappa=\kappa'+i\kappa''$. Likewise, the middle panel
illustrates the corresponding damping, i.e. the relation between frequency $\omega$ and the
imaginary part $\kappa''$ of the Bloch wave vector. Finally, the
right panel shows the density-of-states associated with the dispersion diagram in the left panel. The
difference between the ideal structure ($Q\rightarrow\infty$) and
one employing resonators of finite $Q$ is easily contrasted by comparing the dashed
and solid lines, respectively. We emphasize that the main cause of a finite $Q$ factor
is to smear out van~Hove singularities in the density-of-states and to weaken the slow down of light pulses propagating near the band-edges of the CROW. Of course, the finite $Q$ also introduces damping throughout the entire band, though be most pronounced near the band edges due to slow-light enhanced absorption~\cite{Mortensen:2008}. Finally, we note that quite steep bands appear outside the traditional band of extended states, though of course with a significant attenuation as evident from the middle plot illustrating the $\kappa''$ dependence.

In our previous work we have carefully discussed the influence of the finite $Q$ factor on the saturation of the group index~\cite{Raza:2010,Grgic:2009}. The main conclusion is that in the center of the band the group velocity is (to lowest order in $1/Q$) insensitive to the finite photon life time. On the other hand, at the band edges, initially supporting pronounced slow down, the group velocity scales quite unfavorably with the $Q$ factor, making the slow-light regime challenging to explore.

Here, we focus on the attainable group delay when the slow down and the damping is treated on an equal footing. For the ideal CROW, the group delay is given by $\tau= L/v_g$ . However, in the presence of a finite $Q$, the length $L$ is effectively cut off by the damping length
\textcolor{black} {$\xi=1/2\kappa''$ }associated with the exponential decay in Eq.~(\ref{eq:T}). This gives an upper bound and Eq.~(\ref{eq:taumax}) may in this case formally be rewritten as
\begin{equation}
\tau_{\rm max}(\omega)=\frac{1}{2\kappa''}\frac{\partial \kappa'}{\partial\omega}.
\end{equation}
Combining the full results for $\kappa'$ and $\kappa''$ and expanding in
$1/Q$ we get~\cite{Raza:2010}
\begin{equation}\label{eq:taumax_expansion}
\tau_{\rm max}(\omega) =\tau_p +  {\cal O}(Q^{-1}).
\end{equation}
The main conclusion from this analytical exercise is that the maximal group delay is limited by the photon
life time $\tau_p = Q/\Omega$ of the individual resonators,
independently on the actual frequency. While being a quite intuitive results, it has important and overlooked consequences with respect to how much delay one may envision in future designs of optical buffers and delay-line architectures. Despite the reduced group
velocity near the band edges, the advantage of a slowly advancing
wave package is balanced by a reduced propagation length, see the
middle panel of Fig.~\ref{fig1}. We emphasize that compared to a single resonator, the CROW may of course
offer the advantage of an increased bandwidth. Likewise, the strongly suppressed group-velocity dispersion at the band center might also be beneficial in some applications.

In the following we discuss how this result is modified in the presence of disorder. However, we may at this stage anticipate that disorder may only further limit
time delay, thus in general adding the 'lesser sign' in Eq.~(\ref{eq:taumax<taup}) as compared to the equality derived in Eq.~(\ref{eq:taumax_expansion}).

\subsection{The influence of disorder}

We next turn to disordered waveguides, formally allowing for CROWs composed of resonators with a resonator-to-resonator fluctuation in the resonance frequency $\Omega_j$ and the linewidth $\Gamma_j$ as well as fluctuations in couplings $\gamma_{lj}$ between neighboring resonators. For simplicity we will neglect fluctuations in the linewidth so that all resonators have the same $Q$ factor. For the resonance frequencies and the couplings we will further assume uncorrelated Gaussian distributions $P(\Omega_j)$ and $P(\gamma_{lj})$, respectively. The distributions have mean values $\Omega$ and $\gamma$ corresponding to the ideal CROW parameters, while fluctuations $\sigma_{\Omega}=\sqrt{\delta^2\Omega}$ and $\sigma_{\gamma}=\sqrt{\delta^2\gamma}$ around the mean values can be varied to mimic different strengths of disorder.

We employ a Green's function method to calculate the transmission and the density-of-states which allows us to evaluate Eq.~(\ref{eq:taumax}) for each member of the ensemble. The Green's function is formally obtained by inverting the matrix ${\boldsymbol H}$. However, since its dimensions are formally infinite we imagine a segment of disordered CROW (containing $N$ resonators) sandwiched between to semi-infinite ideal CROWs, i.e. with no disorder. With the aid of Dyson's equation this apparently unsolvable problem can fortunately be turned into a matrix problem of finite dimension (corresponding to the dimension $N$ of the disordered segment) and the
retarded Green functions of the CROW can now be found from a numerical inversion of a sparse $N\times N$ matrix problem~\cite{Datta2000,Datta1995}
\begin{equation}\label{eq:green}
{\boldsymbol {\cal G}}(\omega)=\big[\omega {\boldsymbol I}- {\boldsymbol H}
-{\boldsymbol \Sigma}(\omega)
\big]^{-1},
\end{equation}
where $\boldsymbol I$ is a unit matrix and the couplings to the two semi-infinite ideal CROWs are accounted for by a complex-valued frequency dependent self-energy ${\boldsymbol \Sigma}$. The details of this approach are given in the appendix~\ref{appendix:A}, which lists expressions how to obtain the transmission $T$ and the density-of-states $\rho$ from the Green's function ${\boldsymbol{\cal G}}$.

Once the Green function is obtained, the maximal time delay, Eq.~(\ref{eq:taumax}), is thus easily evaluated with the aid of Eqs.~(\ref{eq:T_Tr}) and (\ref{eq:rho}). We take advantage of standard matrix inversion routines to numerically study the statistical properties of large ensembles of disordered CROWs. In principle this allows us to study statistical moments to any order, but for simplicity we will here focus on the average properties (first moment) and only highlight the CROW-to-CROW fluctuations (second moment) by displaying results for particular members, chosen randomly from the large CROW ensemble.

\begin{figure}[h!]
\begin{center}
\includegraphics[width=0.8\columnwidth]{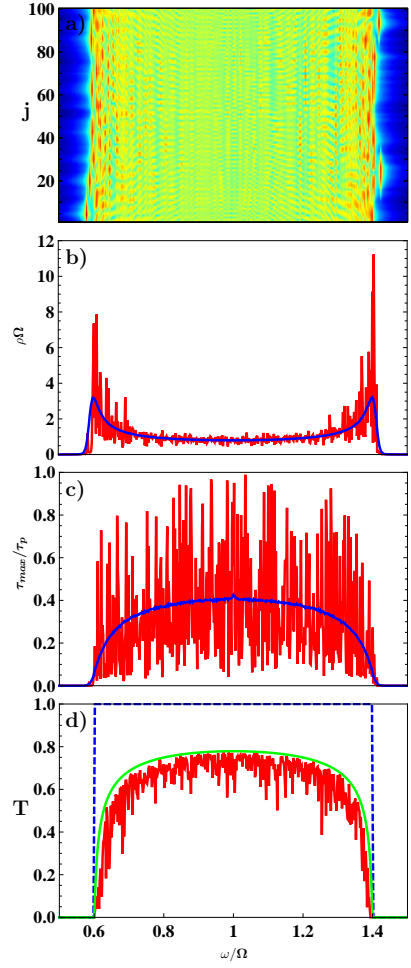}
\end{center}
\caption{Properties of a disordered CROW, with blue lines indicating ensemble-averaged properties while the red lines illustrate the results for a particular realization of the disorder, thus emphasizing pronounced CROW-to-CROW fluctuations. Panel (a) shows the local DOS $\rho_j$ for a particular realization of the disorder and panel (b) shows the corresponding results for the total DOS $\rho$ (per resonator). Panel (c) shows the maximal group delay $\tau_{\rm max}$. Panel (d) shows results for the transmission. For comparison, the dashed line shows the unity transmission for an ideal crow, while the green line is for a non-disordered CROW, but with a finite $Q$. } \label{fig2}
\end{figure}

In the panel b) of the Figure~\ref{fig2} we show results for the ensemble-averaged DOS (blue lines). It is clearly seen how disorder, in addition to a finite $Q$, serves to further broaden the ensemble-averaged DOS near the band edges. Comparing these results to the DOS associated with one particular realization of the disordered CROW (red line), it is however clear that pronounced sample-to-sample fluctuations are to be expected. In particular, the formation of Anderson localized states near the band edges, due to the strong interference of light waves, is apparent. These fluctuations in the DOS are quite naturally inherited by other central quantities, such as the maximal group delay $\tau_{max}$ and the transmission $T$. In panel c) the value of  $\tau_{max}$ is normalized with photon lifetime in single resonator $\tau_p$. For the ensemble-averaged maximal delay time, disorder is seen to further suppress $\tau_{\rm max}$ below the bounds by $\tau_p$. However, from a practical point of view it is alarming to see fluctuation comparable to the mean value, as indicated by the strongly fluctuating results for a particular realization of the disorder (red line). For the transmission in panel d) we see a similar picture with a strong suppression of the transmission near band edges, but with the pronounced transmission fluctuations appearing throughout the entire band. For comparison, the dashed line shows the result of unity transmission ($T=1$) for an ideal CROW with infinite $Q$, while the green line shows the pronounced suppression of transmission in the presence of a finite $Q$, but in the absence of any additional disorder. The quite abrupt drop in transmission near the band edges is associated with slow-light enhanced absorption~\cite{Mortensen:2007} as compared to the center of the band where slow-light enhancement is almost absent.

\begin{figure}[h!]
\begin{center}
\includegraphics[width=\columnwidth]{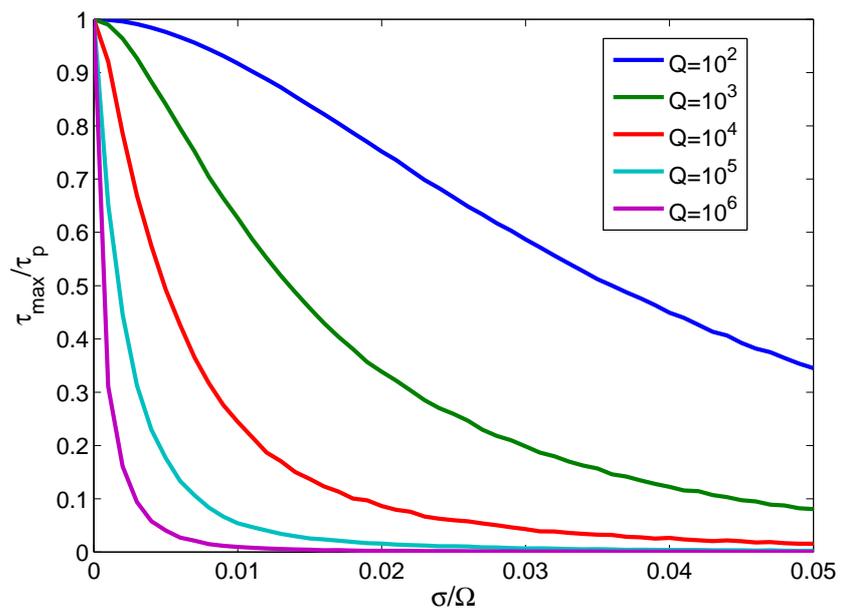}
\end{center}
\caption{Maximal group delay $\tau_{\rm max}$ (at band center) versus disorder strength $\sigma=\sigma_\gamma=\sigma_\Omega$.} \label{fig3}
\end{figure}

There is an interesting interplay between the finite $Q$ factor and the amount of disorder in the structure. In Figure~\ref{fig3} we plot $\tau_{max}/\tau_p$ as a function of disorder strength $\sigma/\Omega=\sigma_\gamma/\Omega=\sigma_\Omega/\Omega$, evaluated for a frequency corresponding to the band center. The different curves represent different values of the $Q$ factor. Quite intuitively, the lower the $Q$ value the less sensitive is the result to disorder, keeping in mind that $\tau_{\rm max}$, $T$, and $n_g$ themselves would be heavily suppressed in the presence of a low $Q$. For higher $Q$ values, it consequently becomes increasingly challenging, in terms of disorder, to take full advantage of the high intrinsic photon life time $\tau_p$.

\section{Conclusion}
\label{sec:conclusion} In conclusion, we have derived an explicit
relation for the dispersion relation of CROWs made from resonators
with a finite $Q$ factor. A finite $Q$ profoundly influences the
van~Hove singularities near the band edges with a resulting
limitation of the group index while at the center of the band the
dispersion properties are less affected. Simple analytical
expressions are supported by calculations of the group velocity,
demonstrating how the $Q$ enters on an equal footing with the
coupling $\gamma$ corresponding to the competing time scales
associated with photon decay and tunneling. In the context of
practical applications involving the group delay, we note that the
maximal attainable group delay appears as a balance between the
reduced group velocity and the the decay length. Explicit
calculations show that irrespectively of the underlying
bandstructure, the maximal group delay is limited by the photon life
time of the resonators. This illustrates the importance of
addressing propagation loss and slow-light on an equal footing. Any inevitable presence of disorder will serve to further suppress the attainable group delay and pronounced sample-to-sample fluctuations may arise.

\section*{Acknowledgments}

This work is financially supported by the Villum
Kann Rasmussen Centre of Excellence NATEC (Nanophotonics
for Terabit Communications).
\newpage
\appendix
\section{Details of Green's function approach}
\label{appendix:A}
The self energy in Eq.~(\ref{eq:green}) is given by ${\boldsymbol \Sigma}={\boldsymbol\Sigma}_{\scriptscriptstyle L}+{\boldsymbol\Sigma}_{\scriptscriptstyle R}$ where the contributions from the left and right semi-infinite CROWS are given by
\begin{equation}
\big\{{\boldsymbol\Sigma}_{\scriptscriptstyle p}(\varepsilon)\big\}_{jl}=
-\gamma \exp\left[i \kappa(\omega) a\right]\,\delta_{js_p}\delta_{s_pl}
\end{equation}
with $s_{\rm L}=1$ and $s_{\rm R}=N$. The wave vector is related to
the energy through the usual cosine dispersion relation derived above, see Eq.~(\ref{eq:dispersion}), corresponding to

\begin{equation}
\exp\left[i \kappa(\omega) a\right]=\frac{\Omega-\omega}{2\gamma}+i\sqrt{1+\frac{(\Omega-\omega)^2}{4\gamma^2}}.
\end{equation}

The transmission probability may now conveniently be written as a trace formula

\begin{equation}\label{eq:T_Tr}
T(\omega)={\rm Tr}
\big[{\boldsymbol\Gamma}_{\rm \scriptscriptstyle L}(\omega)
{\boldsymbol{\cal G}}(\omega)
{\boldsymbol\Gamma}_{\rm \scriptscriptstyle R} (\omega)
{\boldsymbol{\cal G}}^\dagger(\omega) \big],
\end{equation}
where
\begin{equation}
{\boldsymbol\Gamma}_{\scriptscriptstyle p}(\omega)
=i\big[{\boldsymbol\Sigma}_{\scriptscriptstyle p}(\omega) -
  {\boldsymbol\Sigma}_{\scriptscriptstyle p}^\dagger(\omega)\big].
\end{equation}

Likewise, the total density-of-states (per resonator) is given by

\begin{equation}\label{eq:rho}
\rho(\omega)=\frac{1}{N}\sum_{j=1}^N \rho_j(\omega),
\end{equation}
with the corresponding local density-of-states governed by the diagonal part of the Green's function,

\begin{equation}
\rho_j(\omega)=-\frac{1}{\pi}{\rm Im}\{{\cal G}_{jj}(\omega)\}.
\end{equation}

\newpage



\newpage

\end{document}